\documentclass{article}

\usepackage[final]{neurips_2023}

\usepackage[utf8]{inputenc} 
\usepackage[T1]{fontenc}    
\usepackage{hyperref}       
\usepackage{url}            
\usepackage{booktabs}       
\usepackage{amsfonts}       
\usepackage{nicefrac}       
\usepackage{microtype}      
\usepackage{xcolor}         
\usepackage{float}
\usepackage{graphics}
\usepackage{graphicx}

\title{Characteristics of Political Misinformation Over the Past Decade}

\author{%
  Erik J.~Schlicht \\
  Misinformation-Monitor\\
  \texttt{misinfo-monitor.org} \\
  \texttt{erik@misinfo-monitor.org} \\
}

\begin{document}

\maketitle

\begin{abstract}
Although misinformation tends to spread online, it can have serious real-world consequences.  In order to develop automated tools to detect and mitigate the impact of misinformation, researchers must leverage algorithms that can adapt to the modality (text, images and video), the source, and the content of the false information.   However, these characteristics tend to change dynamically across time, making it challenging to develop robust algorithms to fight misinformation spread.  Therefore, this paper uses natural language processing to find common characteristics of political misinformation over a twelve year period.  The results show that misinformation has increased dramatically in recent years and that it has increasingly started to be shared from sources with primary information modalities of text and images (e.g., Facebook and Instagram), although video sharing sources containing misinformation are starting to increase  (e.g., TikTok).  Moreover, it was discovered that statements expressing misinformation contain more negative sentiment than accurate information.  However, the sentiment associated with both accurate and inaccurate information has trended downward, indicating a generally more negative tone in political statements across time.  Finally, recurring misinformation categories were uncovered that occur over multiple years, which may imply that people tend to share inaccurate statements around information they fear or don't understand (Science and Medicine, Crime, Religion), impacts them directly (Policy, Election Integrity, Economic) or Public Figures who are salient in their daily lives.  Together, it is hoped that these insights will assist researchers in developing algorithms that are temporally invariant and capable of detecting and mitigating misinformation across time.

 \end{abstract}

\section{Introduction}

Misinformation is a statement that contains false or misleading information, and can result in serious consequences, including the erosion of civil discourse, political paralysis, uncertainty, in addition to alienation and disengagement (Kavanagh \& Rich, 2018; Hook \& Verdeja, 2022). Despite its serious impact on individuals and society, misinformation is known to be shared more than valid information (Vosoughi, et al, 2018), and the reasons that misinformation is propagated are diverse and include cognitive factors (Del Vicario, et al , 2016; Ecker, et al, 2022), socio-affective factors (Ecker, et al, 2022), incentives (Ceylan, et al, 2023) and changes in the information system (Kavanagh \& Rich, 2018, Chen et al, 2023).

As a result, finding scalable solutions to detect and mitigate the impact of misinformation has proven challenging, although many efforts have demonstrated some promise (Conroy, et al., 2015; ,Aldwairi \& Aldwairi, 2018, RAND, 2018). Part of the challenge facing researchers is that misinformation can be propagated through many modalities (e.g., text, image and video), thereby increasing the algorithmic complexity and computational resources necessary to deploy scalable solutions. Moreover, the sources from which misinformation is spread can effortlessly adapt to any mitigation attempts (e.g., account removal), which can quickly become a real-life version of the Whac-A-Mole game.

In order for robust and effective solutions to be deployed, first we must understand the features of misinformation that are temporally invariant, since the content, sources and modalities associated with misinformation are likely to change across time. Although some researchers have explored temporal patterns of misinformation on social media (Allcott, et al, 2019), they focused on relatively narrow timelines that were less than five years in range. Therefore, this effort explores the trends associated with political misinformation over a twelve year period, leveraging tools from natural language processing to uncover common themes in misinformation across time. By uncovering these insights, it may allow researchers to develop more robust tools to detect and mitigate misinformation, and the next section will detail the data used to this end. 

\section{PolitiFact Data}
\label{data}

In order to investigate trends in political misinformation across time, twelve years of fact-checked \href{https://www.politifact.com/} {PolitiFact} data was obtained, ranging between 2011 through 2023.  PolitiFact is owned by the nonprofit  \href{https://www.poynter.org/about//} {Poyneter Institute for Media} with the objective of improving the relevance, ethical practice and value of journalism.  As part of that objective, PolitiFact verifies the accuracy of claims made throughout various online and media sources.  By using this fact-checked information,  it offers a source of ground-truth for information accuracy with relatively high confidence.  

Approximately sixteen thousand statements were collected across twelve years, and further classified into three categories: 
\begin{itemize}
  \item \textbf{Accurate:}  this information class was given to fact-checked ratings with TRUTH-O-METER scores of  TRUE or MOSTLY-TRUE.
  \item  \textbf{Misinformation:}  this information class was given to fact-checked ratings with TRUTH-O-METER scores of  PANTS-ON-FIRE or FALSE.
  \item  \textbf{Mixed-Accuracy:}  this information class was given to fact-checked ratings with all other TRUTH-O-METER ratings not contained in the classes above.
  \end{itemize}

This statement classification resulted in the following distribution of information accuracy in the sample:

\begin{table}[h]
  \caption{Information Classes Across Data}
  \label{class-table}
  \centering
  \begin{tabular}{ll}
    \toprule
    \multicolumn{2}{c}{PolitiFact Sample}      \\
    \cmidrule(r){1-2}
    Class     & Count    \\
    \midrule
    Accurate Statements & 3,353     \\
    Misinformation     & 7,720  \\
    Mixed-Accuracy Statements     & 4,921       \\
    \textbf{Total}     & \textbf{15,994}       \\
    \bottomrule
  \end{tabular}
\end{table}

Using this data and classification system, the following section will show the trends in misinformation across time.  

\section{Trends in Information Accuracy}
\label{trends}

Figure 1 shows how political information accuracy has changed across time.  It is apparent that political misinformation has increased over time, starting around 2017 (red dashed line), when it first became more frequent than accurate or mixed-accuracy information.  It is interesting to note that this is eleven years after the public launch of Facebook and Twitter (2006), and according to \href{https://www.oberlo.com/statistics/how-many-users-does-facebook-have} {Oberlo.com}, this was a time when Facebook had approximately two billion users.  

\begin{figure}[h]
  \centering
  \includegraphics[width=10cm]{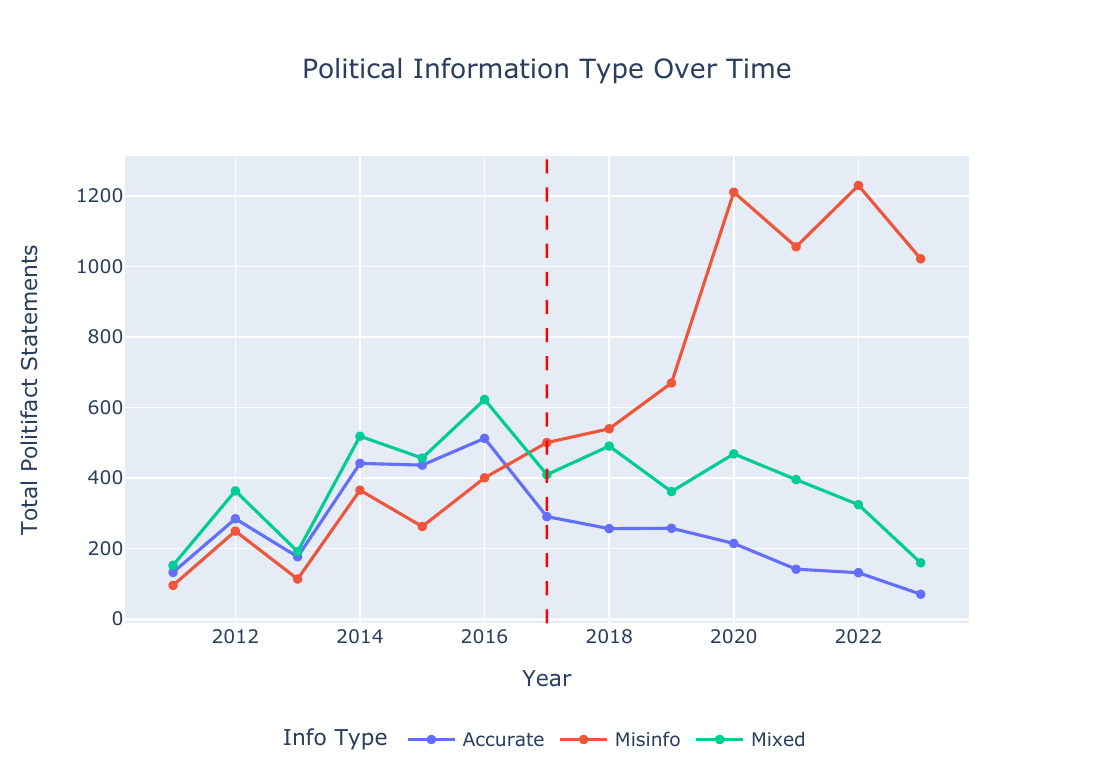}
  \caption{Trends in PolitiFact statement accuracy across time.}
\end{figure}

The approximately ten year window that it took for misinformation to surface on PolitFact following the launch of major social media sites can be due to a number of reasons including the time it takes for misinformation to be noticed by journalists and fact-checkers on these platforms, or even a lack of partnerships that enabled PolitiFact to obtain misinformation data at scale.  

Regardless, it is apparent that the increase in misinformation on PolitiFact is significant, and the next sections will focus on misinformation specifically and explore the primary source modality from which it was shared across time. 

\section{Misinformation Source Modality Trends}
\label{modality}

Since the modality in which misinformation is conveyed (text, images or video) impacts the algorithmic complexity and computational resources necessary to deploy solutions, it is important to understand trends across time.   Figure 2 shows the primary information modality associated with misinformation sources that were responsible for at least five PollitiFact statements for a given year.  For example, Facebook and Twitter (now Meta and X) both have a primary modality of text, since most of the posts shared on these sites are written in text, whereas Instagram and TikTok would be examples of sources with a primary modality of image and video, respectively.  Finally, misinformation statements associated with sources that are entities are labeled with a source modality of individual.

\begin{figure}[h]
  \centering
  \includegraphics[width=10cm]{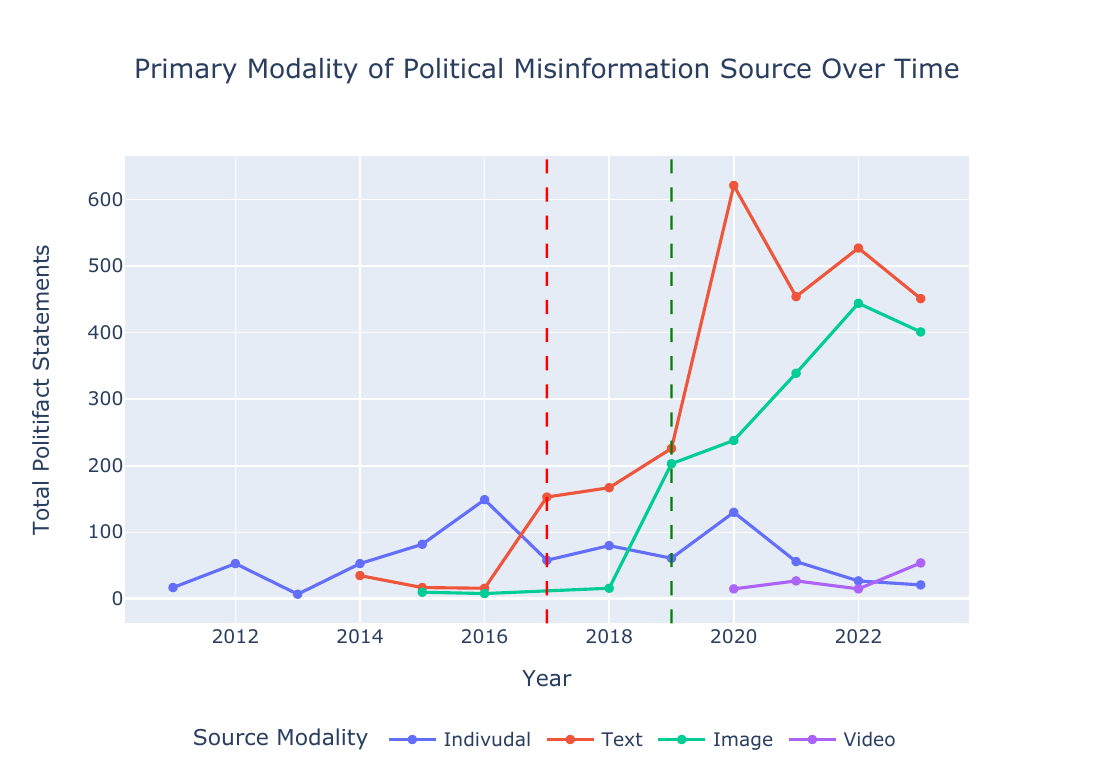}
  \caption{Primary source modality of misinformation across time. }
\end{figure}

It is apparent that misinformation from sources that are primarily in text form increased around 2017, eleven years after the launch of Facebook and Twitter, whereas misinformation from sources associated with images started to increase around 2019, only nine years after the public launch of Instagram.  Moreover, misinformation from sources with a primary modality of video started to appear in 2020.  If misinformation associated with videos follow similar trends as text and images, it would predict an uptick in this modality in the next few years, as TikTok was launched internationally in 2017. 

As a result of the multimodal nature of contemporary misinformation, researchers should no longer rely exclusively on algorithms associated with text (natural language processing) or images/video (computer vision), and transition to multimodality versions available models. Alternatively, they may choose to deploy independent models to detect and mitigate for each modality, but this may require additional complexity.  

Regardless of the solution, these results demonstrate that political misinformation now spans the modality spectrum (Figure 2) and that misinformation has been trending upward in recent years (Figure 1).  Next, tools from natural language processing will be used to explore the sentiment associated with both accurate and inaccurate information across time. 

\section{Information Sentiment Trends}

Previous research found that misinformation relies on emotional content, such as appealing to morality and statements with negative sentiment (Carrasco-Farre, 2022).  Therefore, this section explores the sentiment associated with political statements sampled in this study by leveraging the compound score produced by VADER (Hutto \& Gilbert, 2014).   This score ranges between -1 and +1, with scores less than 0 corresponding to a statement with negative sentiment, scores greater than 0 corresponding to statements with positive sentiment, and scores around 0 corresponding to neutral statements.  

\begin{figure}[h]
  \centering
  \includegraphics[width=10cm]{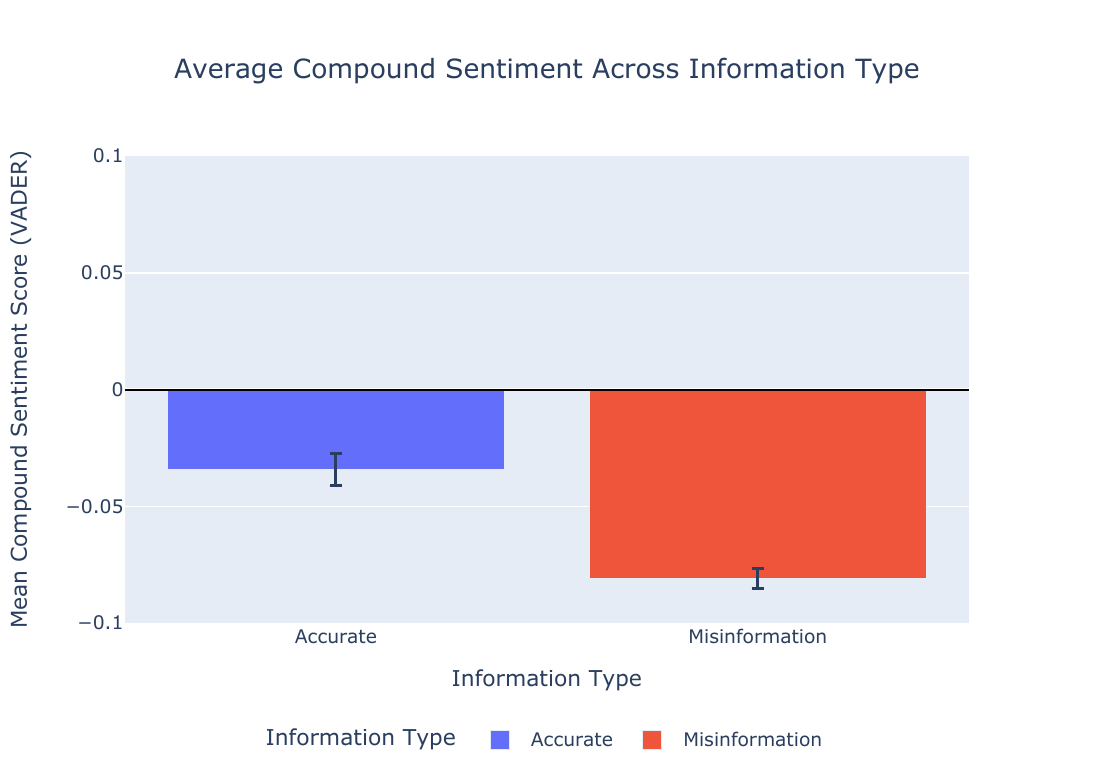}
  \caption{Average statement sentiment for each information type (error bars correspond to \textpm 1 SEM).  }
\end{figure}

By averaging the compound sentiment scores across years (Figure 3), similar trends were realized where misinformation was associated with more negative sentiment (Mean = -.08, SEM = .004) than accurate statements (Mean = -.03, SEM = .007).  A  Mann-Whitney U-test found these differences to be significant (p<.001).  This supports the notion that misinformation relies on emotional content to propagate at higher rates than valid information (e.g., clickbait).

\begin{table}[h]
  \caption{Examples of Statement Sentiment}
  \label{class-table}
  \centering
  \resizebox{\columnwidth}{!}{\begin{tabular}{ll}
    \toprule
    \multicolumn{2}{c}{PolitiFact Statement Sentiment}      \\
    \cmidrule(r){1-2}
    Statement     & VADER Compound Score    \\
    \midrule
 Says Newt Gingrich aligned with Nancy Pelosi on global warming.  & 0.1531   \\
 In New York City, "an entry level janitor gets paid twice as much as an entry level teacher." & 0   \\
 Says it cost Massachusetts taxpayers \$100,000 when Mitt Romney and his staff purchased computer hard drives. & -0.1027    \\
    \bottomrule
  \end{tabular}}
\end{table}

In oder to investigate trends in misinformation across time, average sentiment was calculated separately within each year and information type (Figure 4). The results show a slight negative trend in PolitiFact statements over time for both accurate and inaccurate statements.  This is somewhat surprising, since the objective of accurate information is to relay facts rather than propagate sensational claims. Therefore, this may suggest that accurate information has adopted similar hyperbolic reporting tactics over time as misinformation in order to compete for the attention of readers. 

Despite the fact that the results show significant differences in sentiment between information type, they are very small and do not correspond to large perceptible differences in statements (Table 2).  Therefore, although these trends are interesting, they do not correspond to major tendencies towards negative dialog over years.  

\begin{figure}[h]
  \centering
  \includegraphics[width=10cm]{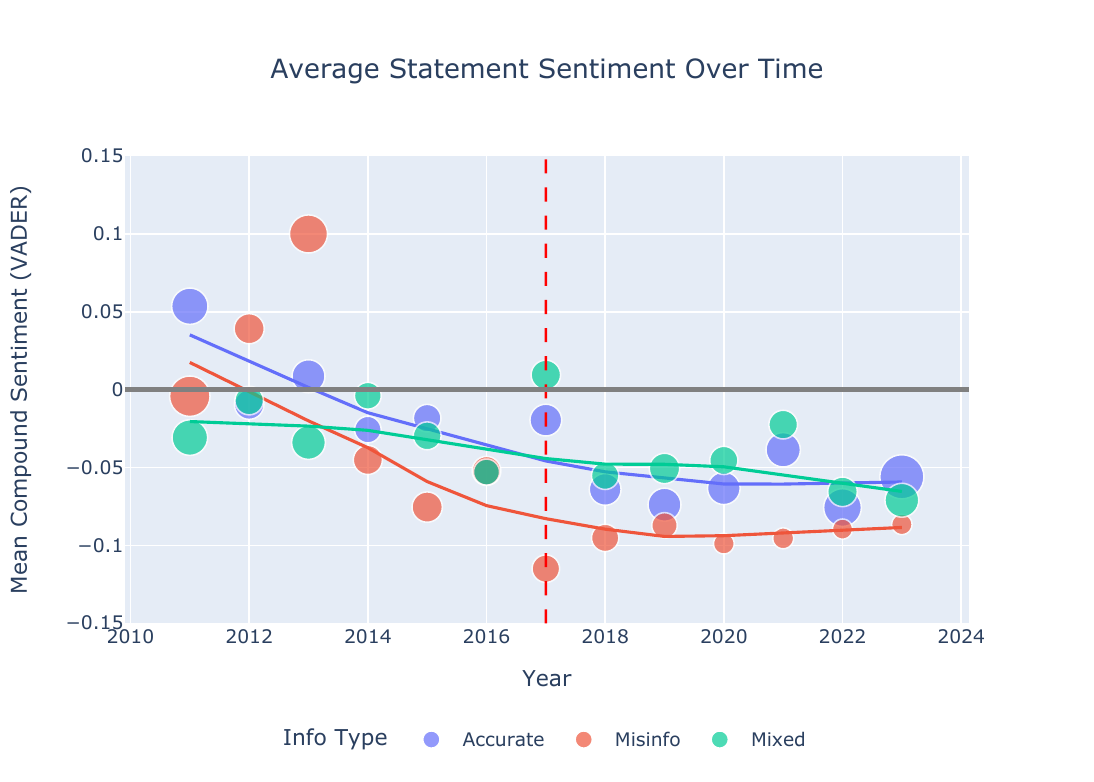}
  \caption{Average statement sentiment for each year (marker size proportional to SEM).  }
\end{figure}

\section{Misinformation Topics Across Years}

Another line of investigation explored in this effort was to use a new technique in natural language processing (BERTopic) to discover topics associated with PolitiFact statements (Grootendorst, 2022).  The model uses document embeddings to cluster into coherent topics in order to uncover themes in text.  Misinformation statements from each year were input into the algorithm, allowing it to converge on the best number and type of clusters for each year independently (Table 3).  

\begin{table}[h]
  \caption{Misinformation Topics Across Years. }
  \label{class-table}
  \centering
  \resizebox{\columnwidth}{!}{\begin{tabular}{lll}
    \toprule
    \multicolumn{2}{c}{PolitiFact Statement Topics}      \\
    \cmidrule(r){1-3}
    Year     & Topic Description (Number of Statements in Topic)  & Number  \\
    \midrule
2023 & Joe Biden (359),  Space (79), COVID (72), Spanish Content (62), Elon Musk (33), Walmart Refund (32) & 11 \\
 & State Politics (31), Wildfires (30), Medicine (15), Celebrities (15), Mass Shootings (14)	 \\
\hline
2022 & COVID (160),  Ukraine War (152), Election Fraud (118), Universal Healthcare (84), Joe Biden (61)  & 21\\
 &  Celebrities (57), Mass Shootings (43), Nancy Pelosi (36), Gas Prices (26), Elon Musk (25), Hillary Clinton (23)  \\
 & Queen Elizabeth (31), Space  (19), Adolf Hitler (18), Monkeypox (16), Climate Change (15),     \\
 & Alkaline Foods (14), Defund Police (14), Electric Vehicles (12), Baby Formula (11), Corporate Conspiracy (10) \\
\hline
2021 & COVID (359),  Election Fraud (94), Joe Biden (58),  Hillary Clinton (55), Capitol Riot (54) & 17\\
 & Taxes (43),  Meghan Markle (23), Afghanistan (20),  Gas Prices (20), Climate Change (20), Immigration (19),\\
 &  Nancy Pelosi (17), Facebook (14),  George Floyd (13), Mass Shootings (13), Electric Vehicles (13), Racism (10)\\
\hline
2020 & Joe Biden (417),  COVID (300),  BLM (89), PPE Masks (29),  Mass Shootings (26), Barack Obama (20) & 9\\
 &  Iranian Rockets (16),  Kobe Bryant (15), Gretchen Whitmer (10)\\
\hline
2019 & Donald Trump (96),  Political Quotes (94),  State Politics (58), Joe Biden (28),  Border Wall (27) & 11\\
 & Immigration (21),  Vaccines (20), Alexandria Ocasio-Cortez (18), Abortion (18), Democratic Policy (12), Muslims (12)\\
\hline
2018 & Democratic Policy (118),  Barack Obama (65),  Donald Trump (25), Second Amendment (18) & 7\\
&  Immigration (18), Mass Shootings (14), Airline (11)  \\
\hline
2017 & Barack Obama (110),  Universal Healthcare (27), Republican Policy (27), State Politics (24), Muslims (22) & 11\\
&  Space (21), Mass Shootings (20), Hurricanes (16), Celebrities (15),  Donald Trump (13), Abortion (12) \\
\hline
2016 & Hillary Clinton (111),  Taxes (88), Barack Obama (55), Abortion (13) & 4\\
\hline
2015 & Democratic Politicians (224), Abortion (13) & 2\\
\hline
2014 & Universal Healthcare (49), Barack Obama (48), Taxes (29), Scott Walker (24),  Economic Disparity  (18) & 6\\
&  Mass Shootings (11)\\
\hline
2013 & Taxes (36), State Politics (48), Scott Walker (11)  & 3\\
\hline
2012 & Universal Healthcare (35), Scott Walker (31),  Barack Obama (17), Taxes (16), State Politics (13) & 6\\
&  Mitt Romney (13)\\
\hline
2011 & Barack Obama (20), State Politics (13)  & 2\\
    \bottomrule
  \end{tabular}}
\end{table}

Subjective descriptions were provided for each cluster/topic to better understand common topics across years.  Table 3 shows each of these topic descriptions with the number of misinformation statements that are associated with each topic. The number of topics BERTopic converged on increased across time, suggesting that not only is there more misinformation in recent years (Figure 1), but that it spans a greater number of topics (Table 3).  Note that the topic description corresponds to the primary theme or entity associated with each cluster.  In the case of entities, it does not discriminate between those where the entity was the target of misinformation or if they were responsible for stating the misinformation.  

Exploring topics within each year, it is apparent that some topics are heavily aligned with current events and themes that are temporally localized (e.g., Capitol Riot). However, there are some topics that recur across years and the next section will explore those topics further.  

\section{Recurring Misinformation Topics}

In order to facilitate insights into topics that occur across multiple years, Figure 5 depicts recurring topics and the number of statements that correspond to each recurring topic.  The number of years topics recur ranges between two and seven years, with Mass Shootings and Barack Obama occurring most frequently across years. However, when focusing on the total number of statements that correspond to each recurring topic, Joe Biden and COVID are the topics that correspond to the greatest number of misinformation statements. 

\begin{figure}[h]
  \centering
  \includegraphics[width=10cm]{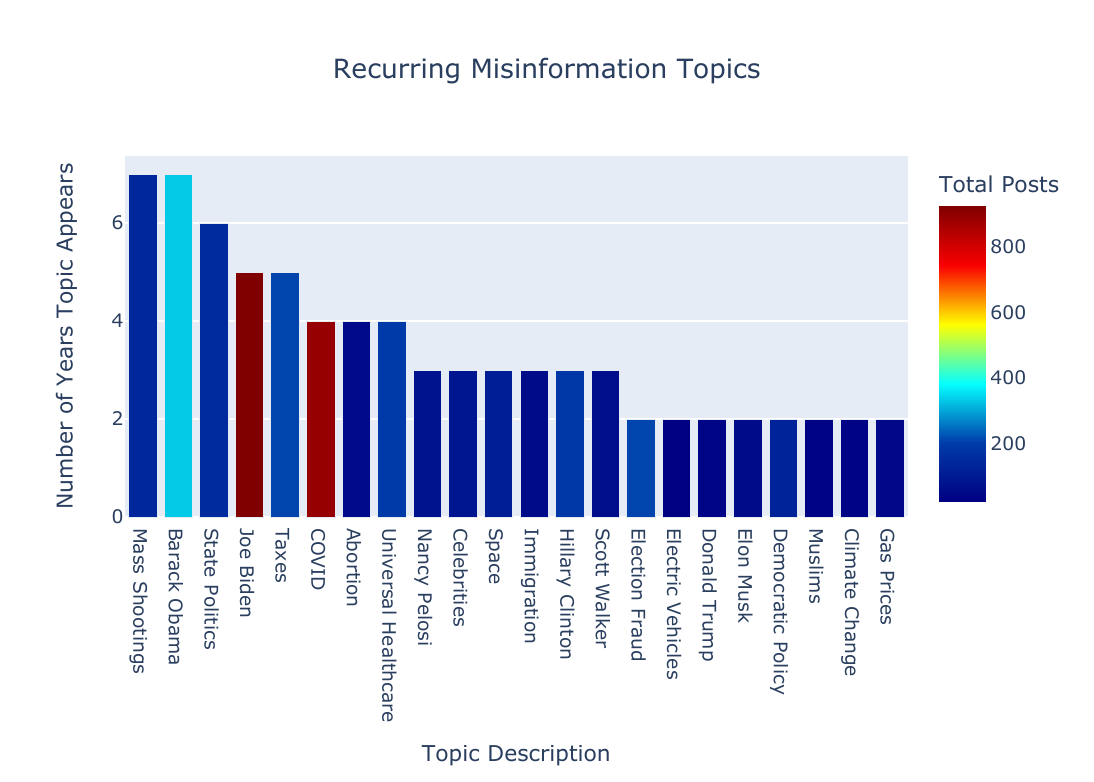}
  \caption{Recurring misinformation topics. }
\end{figure}

\begin{table}[h]
  \caption{Examples of Statements for Recurring Misinformation Topic Categories, }
  \label{class-table}
  \centering
  \resizebox{\columnwidth}{!}{\begin{tabular}{llp{11cm}}
    \toprule
    \multicolumn{3}{c}{Recurring Misinformation Topic Categories}      \\
    \cmidrule(r){1-3}
   Category     & Topic Description & Example Statement  \\
    \midrule
Public Figures & Joe Biden, & Joe Biden just resigned from the White House. \\
  & Barack Obama  & Says President Barack Obama is a socialist. \\
 & Hillary Clinton & Navy SEALs arrested Hillary Clinton.  \\
& Nancy Pelosi  & Says Nancy Pelosi has been executed. \\
 & Scott Walker  & Says Gov. Scott Walker dropped or was kicked out of college short of a degree. \\
 & Elon Musk  & Elon Musk is dead.  \\
 & Celebrities  & U.S. comedian Kevin Hart in critical condition after gory car crash. \\
& Donald Trump  & Says Shaquille O’Neal said Donald Trump is possibly the best president. \\
\hline
Science and Medicine & COVID & COVID-19 vaccines have killed 676,000 Americans. \\
& Space & NASA faked footage of astronauts on the International Space Station. \\
 & Abortion & Almost 95 \% of all (Planned Parenthood) pregnancy services were abortions. \\
 & Climate Change & Says Greta Thunberg said, the climate crisis doesn’t exist. \\
 & Electric Vehicles & When electric cars get in accidents,  they explode, they catch fire very very badly because of the lithium batteries \\
 \hline
Policy & Taxes & The yearly cost of religious tax exemptions: \$71 billion. If the church paid taxes, everyone would only have to pay 3\% taxes.\\
& Universal Healthcare & President Obama's health care law is a government takeover of healthcare. \\
& State Politics & Campaign finance disclosures showed Fulton County District Attorney Fani Willis took part in a massive money laundering and election fraud scheme. \\
& Democratic Policy & Voting for any Democrat gets you socialism, undefended open borders, immediate tax increases, 100\% government-run health care.\\
& Immigration & 2,000 illegal aliens were arrested by ICE in 2017 for murders committed here in the United States! \\
 \hline
 Election Integrity & Election Fraud & Fluctuating Georgia U.S. Senate runoff vote count is evidence of election fraud. \\
  \hline
  Crime & Mass Shootings & The mass shooting in Highland Park, Illinois, was a false flag. \\
  \hline
 Economic	 & Gas Prices & In April 1997, there was a gas out conducted nationwide in protest of gas prices. Gasoline prices dropped 30 cents a gallon overnight. \\
 \hline
Religion & Muslims & Says Islamic studies professor Tariq Ramadan said Muslims are here to colonize the U.S. and Canada and spread Sharia law and won’t hesitate to use violent Jihad if they have to. \\
    \bottomrule
  \end{tabular}}
\end{table}

Taking this aggregation a level higher, subjective categories were provided across topics in an attempt to understand the high-level categories of recurring misinformation topics (Table 4).  The categories span seven separate areas with Public Figures, Science and Medicine, and Policy containing the most recurring topics, whereas other topic categories only contain one recurring topic within each. 

\section{Conclusions}

Overall, these results suggest that the advent of social media has allowed misinformation to propagate more rapidly, spanning a greater number of topics and modalities (Figures 1 \& 2, Table 3).  Moreover, the sentiment results (Figure 3) support the idea that misinformation appeals to people's emotions through negative claims (Grootendorst, 2022) that are often shared at higher rates than accurate information (Vosoughi, et al, 2018).  What's more, the sentiment trends across time show that accurate information has also adopted a slightly more negative tone, which may imply that valid information is starting to be portrayed in an emotional context to complete for the attention of readers (Figure 4).   

This paper also uncovered several recurring topics of misinformation (Figure 5) spanning seven categories (Table 4), which provides insight into temporally invariant themes of misinformation.  Taken in the context that misinformation is emotionally charged, these categories could suggest that people tend to share inaccurate statements around information they fear or don't understand (Science and Medicine, Crime, Religion), that impacts them directly (Policy, Election Integrity, Economic) or Public Figures who are salient in their daily lives.  

The impact of misinformation on our democracy has been well studied and includes the erosion of civil discourse, political paralysis, uncertainty, in addition to alienation and disengagement (Kavanagh \& Rich, 2018; Hook \& Verdeja, 2022).  A necessary first condition to mitigating these outcomes is to reduce the availability of misinformation.   Since social media provides the opportunity for misinformation to be spread rapidly, automated methods for detecting and removing inaccurate misinformation is one approach to accomplish this objective.  

As a result, it is hoped that these insights will help researchers and engineers to develop robust algorithms to detect and mitigate misinformation. Whatever the solution, these algorithms will need to span multiple modalities and capture that major topic categories that correspond to misinformation that were uncovered in this investigation.

\section*{References}
{
\small

Aldwairi, M. and Aldwairi, A. (2018). Detecting fake news in social media networks. \textit{Procedia Computer Science}, 141, 215-222.

Allcott, H., Gentzkow, M. and Yu, C. (2019). Trends in the diffusion of misinformation on social media. \textit{Research and Politics}, 10, 1-8.

Carrasco-Farré, C. The fingerprints of misinformation: how deceptive content differs from reliable sources in terms of cognitive effort and appeal to emotions.  \textit{Humanities and Social Sciences Communications}, 162 (2022). https://doi.org/10.1057/s41599-022-01174-9.

Ceylan, G., Anderson, I. and Wood, W. (2023). Sharing of misinformation is habitual, not just lazy or biased/ \textit{PNAS}, 120, https://doi.org/10.1073/pnas.2216614120.

Chen, S., Lu, X., and Akit, K (2023). Spread of misinformation on social media: What contributes to it and how to combat it. \textit{Computers in Human Behavior}, 141,  \url{https://doi.org/10.1016/j.chb.2022.107643}

Conroy, N.J., Rubin, V., and Chen, Y. (2015). Automatic deception detection: methods for finding fake news. \textit{ASIST 2015: Proceedings of the 78th ASIST Annual Meeting: Information Science with Impact: Research in and for the Community}, 82, 1-4.

Del Vicario, et al. (2016). The spreading of misinformation online. \textit{PNAS}, 113,  \url{https://doi.org/10.1073/pnas.1517441113}.

Ecker, U.K.H., et al. (2022). The psychological drivers of misinformation belief and its resistance to correction. \textit{Nature Reviews Psychology}, 1, 13-29.

Grootendorst, M. (2022). BERTopic: Neural topic modeling with a class-based TF-IDF procedure. \textit{arXiv: Computation and Language}, \url{https://doi.org/10.48550/arXiv.2203.05794}  

Hook, K., Verdeja, E. (2022). Social media misinformation and the prevention of political instability and mass atrocities. \textit{Stimson: Human Security \& Governance}.

Hutto, C., and Gilbert, E. (2014). VADER: A Parsimonious Rule-Based Model for Sentiment Analysis of Social Media Text. \textit{Proceedings of the International AAAI Conference on Web and Social Media}, 8(1), 216-225. 

Kavanagh, J., and Rich, M.D. (2018). Truth Decay: An initial exploration of the diminishing role of facts and analysis in American public life. \textit{RAND Corporation}. ISBN: 978-0-8330-9994-5.

RAND Corporation Website (n.d.). Tools that fight disinformation online.  \url{https://www.rand.org/research/projects/truth-decay/fighting-disinformation/search.html}

Vosoughi, S., Roy, D., and Aral, S. (2018). The spread of true and false news online. \textit{Science}, 359, 1146-1151.

}


\end{document}